\documentclass[printer,amssymb,amsmath]{tMOP2e}
\usepackage{bm}% bold math

\newcommand{\be}{\begin{equation}}
\newcommand{\ee}{\end{equation}}
\newcommand{\bea}{\begin{eqnarray}}
\newcommand{\eea}{\end{eqnarray}}
\newcommand{\ket}[1]{| #1 \rangle}

\newcommand{\medial}[1]{\langle #1 \rangle}

\newcommand{\tfrac}[2]{\textstyle\frac{#1}{#2}}
\newcommand{\dfrac}[2]{\displaystyle\frac{#1}{#2}}

 \received{February 01, 2006}
 \jname{} %\jvol{} \jnum{}
 \jyear{2006}
 \jmonth{}

\begin{document}

\title{{\itshape A cavity-QED scheme for Heisenberg-limited
interferometry}}

%\title{A cavity-QED scheme for Heisenberg-limited interferometry}

 \author{ DAVID VITALI$^\dag$, STEFAN KUHR$^\ddag$,
 MICHEL BRUNE$^\ddag$, JEAN-MICHEL RAIMOND$^\ddag$\\
 \vspace{6 pt}
 $^\dag$Dipartimento di Fisica, Universit\`a di Camerino,\\
 via Madonna delle Carceri, I-62032 Camerino (MC), Italy\\
 \vspace{4 pt}
 $^\ddag$Laboratoire Kastler Brossel, D\'epartement de Physique
 de l'Ecole Normale Sup\'erieure, \\
 24 rue Lhomond, F-75231 Paris Cedex 05, France}

\maketitle
%David Vitali, Stefan Kuhr, Michel Brune, Jean-Michel Raimond
\date{\today}

\begin{abstract}
We propose a Ramsey interferometry experiment using an entangled
state of $N$ atoms to reach the Heisenberg limit for the
estimation of an atomic phase shift if the atom number parity is
perfectly determined. In a more realistic situation, due to
statistical fluctuations of the atom source and the finite
detection efficiency, the parity is unknown. We then achieve about
half the Heisenberg limit. The scheme involves an ensemble of
circular Rydberg atoms which dispersively interact successively
with two initially empty microwave cavities. The scheme does not
require very high-Q cavities. An experimental realization with
about ten entangled Rydberg atoms is achievable with state of art
apparatuses.
\end{abstract}

%\pacs{42.50.St, 42.50.Pq, 03.65.–w, 39.30.+w}

\section{Introduction}

Quantum limits to noise in spectroscopy \cite{spec,wine94} and
interferometry \cite{mzi,yurkesu2,kitaueda,gerry} have attracted
an increasing interest in the last years. The precision of an
interferometric phase measurement is ultimately limited by the
``Heisenberg uncertainty relation'' between energy and time
\cite{wine96,wine04,wine05,vitt05}. However, this limit can only be reached when $N$
entangled particles are used. This represents a major experimental
difficulty which has been overcome only very recently.  In
Ref.~\cite{wine04} three beryllium ions, and more recently in Ref.~\cite{wine05}
six beryllium ions have been prepared in a
maximally entangled state and used in a Ramsey spectroscopy
experiment achieving a sensitivity close to the Heisenberg limit.
In Refs.~\cite{stei04,zei04} experiments involving respectively
three and four maximally entangled photons are reported, clearly
showing interference fringes three and four times narrower than
those obtained in an experiment with uncorrelated photons. Here,
we propose an easy to implement cavity scheme, involving
significantly more particles ($N > 10$).

The aim of a typical interferometry experiment is the measurement
of the relative phase $\varphi$ between two quantum states
$\ket{g}$ and $\ket{e}$ of a two level atom. The standard approach
is the Ramsey interferometry technique \cite{ramsey}, in which a
first $\pi/2$ pulse produces a superposition of $\ket{g}$ and
$\ket{e}$. After a waiting time, the relative phase is probed with
a second $\pi/2$ pulse. The Ramsey setup is formally equivalent to
an optical Mach-Zehnder interferometer, with the two $\pi/2$
Ramsey pulses corresponding to the input and output $50-50$ beam
splitters. The atomic phase shift corresponds to the differential
phase shift between the two arms of the interferometer
\cite{wine94,yurkesu2}. The uncertainty $\Delta \varphi$ of this
phase measurement is limited by counting statistics to $\Delta
\varphi=1/\sqrt{N}$, where $N$ is the number of detected particles
(atoms for Ramsey spectroscopy, photons in the Mach-Zehnder
interferometer) \cite{wine96}.

This limit holds, however, only if the $N$ particles are
uncorrelated, i.\,e.~in a separable quantum state. If, instead,
the $N$ particles are quantum correlated and share some
entanglement (which is equivalent to say that they are
spin-squeezed \cite{kitaueda,berman}), quantum noise sets a
fundamental limit  to the phase uncertainty, $\Delta \varphi =
1/N$. This ultimate limit is called the Heisenberg limit and is
achieved when a maximally entangled state of the $N$ particles is
used and appropriately measured \cite{wine96,vitt05}.

We present here a variant of a Ramsey interferometry experiment
which approaches the ultimate Heisenberg-limited resolution and
which is realizable with present technology. It involves a
collective interaction of $N$ Rydberg atoms with two high-Q
microwave cavities. In the limit of a large atom-cavity detuning
(dispersive interaction), the $N$ atoms are prepared in a
maximally entangled state, i.\,e.~in an atomic Schr\"odinger cat
state after their collective interaction with the first cavity \cite{agarw97}. This
state is sensitive to a Stark phase shift $\varphi$ applied
between the two cavities. The interaction with the second cavity
reads out this phase shift. The final measurement of the atomic
population reveals interference fringes $N$ times narrower than
those achievable using non-entangled atoms.

The outline of the paper is as follows. In Sec.~\ref{sec:scheme}
the scheme of the experiment is described in detail. In
Sec.~\ref{sec:averaging} we calculate the interference signal
resulting from averaging over the fluctuations of the atom number.
In Sec.~\ref{sec:inversion} we present a modified scheme, which
improves this averaged signal. Finally, in
Sec.~\ref{sec:conditioning} a conditional detection scheme further
improving the sensitivity of the detection of the phase shift is
illustrated, while Sec.~\ref{sec:conclusions} contains concluding
remarks.

\section{Scheme of the interferometry experiment} \label{sec:scheme}

\begin{figure}[!b]
\begin{center}
\includegraphics[width=0.6\textwidth]{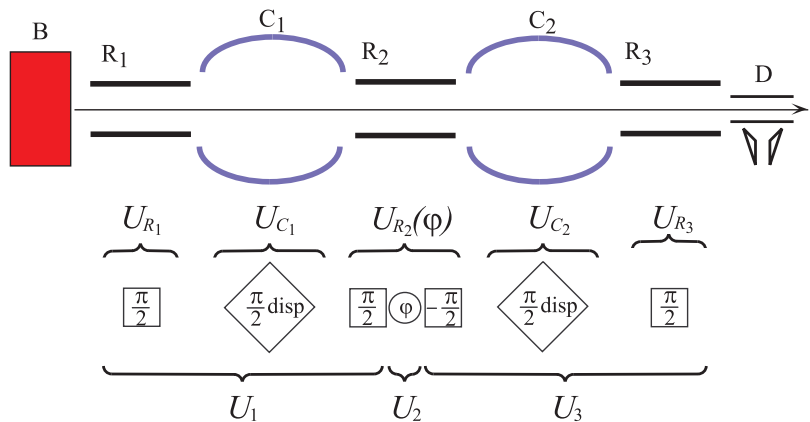}
\end{center}
\caption{\label{fig:scheme} Schematic description of the proposed
cavity-QED interferometry experiment. The $N$ atoms are prepared
in the circular Rydberg state $|g\rangle$ in $B$, then cross the
three classical field zones $R_j$ and the two high-Q cavities
$C_j$, and are finally detected in $D$. The squares denote $\pi/2$
Rabi pulses in $R_j$, the circle denotes the tunable Stark shift
pulse in $R_2$, and the diamonds denote the dispersive atom-cavity
interaction in $C_j$.}
\end{figure}

A schematic description of the proposed interferometry experiment
is shown in Fig.~\ref{fig:scheme}. It involves a collection of $N$
two-level atoms simultaneously crossing a set of microwave
cavities. The two relevant atomic levels are circular Rubidium
Rydberg states $|e\rangle$ and $|g\rangle$ with principal quantum
numbers 51 and 50, and a transition frequency
$\omega_{eg}/2\pi=51$~GHz. The circular Rydberg states are
characterized by a very long lifetime ($\sim 30$~ms) and a very
large dipole moment \cite{rmp}. The atoms are simultaneously
prepared in zone $B$ by the excitation of a velocity-selected
atomic beam effusing from an oven. The $N$ atoms sample, whose
size is negligible at the scale of the millimeter wavelength
$\lambda=2\pi c/\omega_{eg}$, crosses the arrangement of microwave
cavities at the same velocity $v$. Three ``Ramsey zones'', $R_j$
($j=1,2,3$), are separated by two identical microwave Fabry-Perot
cavities $C_1$ and $C_2$, which have a high quality factor $Q$.
The three Ramsey zones are microwave cavities with low $Q$, in
which the Rydberg atoms can be subjected to classical resonant
pulses, generated by standard phase-locked microwave sources. The
three cavities $R_j$ are resonant with the $g\leftrightarrow e$
transition, while the high-Q cavities $C_1$ and $C_2$ are
off-resonant. The atom-cavity detunings are $\delta_{i} =
\omega_{eg} -\omega_{{\rm c}, i}$ ($\omega_{{\rm c}, i}$ is the
mode angular frequency of cavity $i$). The atoms are finally
detected in $D$ by a state selective field-ionization detector,
able to count both the number of atoms in $\ket{e}$, $N_e$ and in
$\ket{g}$, $N_g$.

The $N$ atoms collectively interact with the quantized radiation
mode of $C_1$ and $C_2$, where they are symmetrically coupled
\cite{Dicke}. The interaction of the atomic system with cavity
$i$, in a frame rotating at $\omega_{eg}$, is described by the
Tavis-Cummings Hamiltonian \cite{Tav-Cumm}
\begin{equation}\label{tavcumm1}
\hat H_{I,i}(t) = -\hbar \delta_i \hat a_i^{\dagger}\hat
a_i+i\hbar \frac{\Omega_i(t)}{2} \left(\hat a_i^{\dagger}\hat J_-
- \hat a_i \hat J_+\right),
\end{equation}
where we have introduced the angular momentum Dicke operators
$\hat J_-=\sum_i^N |g\rangle_i \langle e |$, $\hat J_+ = (\hat
J_-)^{\dagger}$ and $\hat J_z=(1/2)\sum_i |e\rangle_i \langle e
|-|g\rangle_i \langle g |$ \cite{Dicke}. In Eq.~(\ref{tavcumm1}),
$\hat a_i$ is the annihilation operator of the quasi-resonant mode
in $C_i$, and $\Omega_i(t)=\Omega_0 \exp\left[-v^2t^2/w^2\right]$
is the vacuum Rabi frequency describing the atom-field
interaction. The time dependency is due to the Gaussian profile of
the radiation mode with waist $w$ by the moving atomic sample.

We are interested in the dispersive limit in which the two
cavities $C_1$ and $C_2$ are detuned far from the atomic
transition, i.e., $\delta \gg \Omega_0 \sqrt{N},\gamma_{\rm c}$
($\gamma_{\rm c}$ is the cavity decay rate).
%As shown in Ref.~\cite{agarw97}, in this limit the dynamics, even
%in the presence of cavity damping, is well approximated by the
%Hamiltonian
%\begin{equation}\label{disp1}
%  \hat H_{{\rm disp}, i}(t) = \hbar \frac{\Omega_i(t)^2}{4\delta_i} \left(\hat J_+ \hat J_- +2 \bar{n}\hat J_z\right),
%\end{equation}
%where $\bar{n}$ is the mean thermal photon number in the cavity.
Using cryogenic techniques together with appropriate ``cooling''
atoms \cite{rmp}, one can approach the zero temperature limit and
therefore we shall assume a mean thermal photon number of zero
from now on. Neglecting the trivial constant of motion $\hat J^2 =
(N/2)(N/2+1)$, Eq.~(\ref{tavcumm1}) is approximated as
\begin{equation}\label{disp2}
  \hat H_{{\rm disp},i}(t) = -\hbar \frac{\Omega_i(t)^2}{4\delta_i} \left(\hat J_z^2-\hat J_z\right),
\end{equation}
explicitly showing that, in the dispersive limit, the atomic
excitation $\hat J_z$ and the photon excitation $\hat
a^{\dagger}_i a_i$, independently, are constants of motion. The
term $\hat J_z^2$ in Eq.~(\ref{disp2}) is responsible for the
generation of atomic entanglement and spin squeezing
\cite{kitaueda}. This is reminiscent of the optical Kerr effect,
which is characterized by a quadratic term in the photon number.
In the atomic as well as in the optical case, this quadratic term
is able, for particular values of the interaction time, to
generate ``Schr\"odinger cats'', i.e., quantum superpositions of
coherent states with different classical phases
\cite{agarw97,molmer99,yurke}.

We adopt the compact representation of an atomic coherent state
with orientation on a Bloch sphere $(\theta,\phi)$, ($0\leq \theta
\leq \pi$, $0\leq \phi < 2\pi$), introduced in \cite{arecchi},
\begin{eqnarray}
|\theta, \phi\rangle &=& e^{-i\theta\left(\hat J_x \sin\phi-\hat J_y\cos\phi\right)} |J,-J\rangle \label{coh}\nonumber \\
&=& \sum_{m=-J}^{J}\left(\begin{array}{c}
  2J \\
  J+m
\end{array}\right) \left(\sin\frac{\theta}{2}\right)^{J+m}\left(\cos\frac{\theta}{2}\right)^{J-m} e^{-i\phi(J+m)}|J,m\rangle,
\end{eqnarray}
where the so-called Dicke states $\ket{J, m}$ are the eigenstates
of $\hat J_z$, and $\hat J^2$ with respective eigenvalues $m$ and
$J(J+1)$.

\begin{figure*}[ptbh]
\begin{center}
\includegraphics[width=0.9\textwidth]{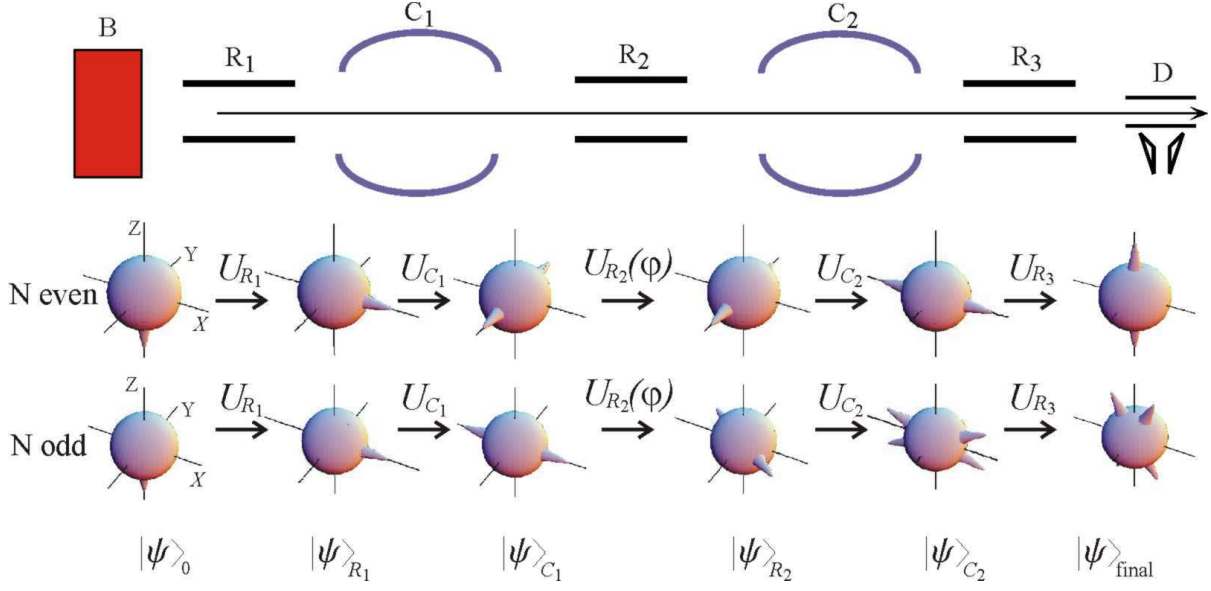}
\end{center}
\caption{\label{fig:blochSperes} Schematic description of the
state of the $N$ Rydberg atoms on a Bloch sphere at the various
stages of the interferometry experiment, both for even $N$ and for
odd $N$. For both parities and at every stage, the $N$ atoms are
in a superposition of atomic coherent states (see
Eqs.~(\protect\ref{cat1ev})-(\protect\ref{cat4ev}) and
Eqs.~(\protect\ref{cat2od})-(\protect\ref{cat4od}) for the
explicit expression of the atomic states in the two cases).}
\end{figure*}

The atoms are initially prepared in the atomic coherent state
$|\psi\rangle _0 =\prod_i |g\rangle_i=|0,0\rangle$ and are then
subjected to a sequence of unitary operations $U_{R_i}$ ($i=1,2,3$)
and $U_{C_j}$ ($j=1,2$) to generate the final state
$|\psi\rangle_{\rm final}$ (see Figs.~\ref{fig:scheme} and
\ref{fig:blochSperes}):
\begin{equation}
 |\psi\rangle_{\rm final}
  = U_{R_3}U_{ C_2}U_{ R_2}(\varphi)U_{C_1}U_{R_1}
  |\psi\rangle_{\rm 0}.
\end{equation}

As a first step, the atoms undergo in $R_1$ a $\pi/2$ pulse,
described by the unitary operator
 $U_{R_1} =\exp{\{i \pi \hat J_y /2\}}$.
It aligns the collective Bloch vector along the $x$-axis,
generating the atomic coherent state
\begin{equation}\label{psi_R1}
|\psi\rangle _{R_1}= U_{R_1}|\psi\rangle_0= | \pi/2,0\rangle.
\end{equation}
The interaction with classical microwave pulses only changes the
orientation of an atomic coherent state on the Bloch sphere
\cite{arecchi} but the $N$ atoms remain disentangled. The atomic
entanglement is produced in $C_1$ by dispersive interaction with
the radiation mode, which is initially prepared in the vacuum
state. As shown in \cite{agarw97}, for specific values of the
interaction time, an initial atomic coherent state
$|\theta,\phi\rangle$ is transformed into a superposition of a
finite number of coherent states with the same $\theta$ and
equally spaced $\phi$ values. An atomic Schr\"odinger cat state
with {\it two} components is the maximally entangled $N$-partite
GHZ state corresponding to the superposition of two atomic
coherent states with opposite orientations on the Bloch sphere,
$\left[|\theta,\phi \rangle + e^{i\beta} |\pi-\theta,\phi +\pi
\rangle\right]/\sqrt{2}$. In our proposal we obtain such an atomic
cat state when \cite{agarw97}
\begin{equation}\label{piphase}
  \int \!\! dt \frac{\Omega_1(t)^2}{4\delta_1}=\frac{\pi}{2}.
\end{equation}
When this condition is satisfied, the interaction in $C_1$ is
represented by the unitary operator
\begin{equation}
  U_{C_1}=\exp\left[i\frac{\pi}{2}(\hat J_z^2-\hat J_z)\right],
\end{equation}
and the state $|\psi\rangle _{C_1}=U_{C_1}|\psi\rangle_{R_1}$ at
the exit of $C_1$ becomes \cite{agarw97}
\begin{equation}
|\psi\rangle_{C_1}^{\rm even} =
 \frac{1}{\sqrt{2}}
 \left[e^{i\frac{\pi}{4}}\ket{\tfrac{\pi}{2},\tfrac{\pi}{2}}+
 (-1)^{\frac{N}{2}}e^{-i\frac{\pi}{4}}
 |\tfrac{\pi}{2},\tfrac{3\pi}{2}\rangle\right], \label{cat1ev}
\end{equation}
if $N$ is even, and
\begin{equation}
|\psi\rangle_{C_1}^{\rm odd}
=\frac{1}{\sqrt{2}}\left[e^{i\frac{\pi}{4}}|\tfrac{\pi}{2},\pi\rangle+
(-1)^{\frac{N-1}{2}}e^{-i\frac{\pi}{4}}|\tfrac{\pi}{2},0\rangle\right],
\label{cat1od}
\end{equation}
if $N$ is odd. This means that we have a cat state directed along
the $y$-axis for even $N$ and along the $x$-axis for odd $N$ (see
Fig.~\ref{fig:blochSperes}).

The goal of the interferometry experiment is to detect a variable
phase difference $\varphi$  between the two components of the cat
state which is applied in $R_2$ by a Stark pulse. This pulse
induce a dephasing between the states $\ket{g}$ and $\ket{e}$,
aligned along the $z$-axis. Thus, we have to rotate the cat state,
aligned along the $y$-axis, to the $z$-axis prior to the
application of the Stark pulse and then rotate it back into its
initial direction. For the case of even $N$, this corresponds to
the pulse sequence
\begin{equation}
  U_{R_2}(\varphi)
   = e^{i\pi \hat J_x/2}
     e^{i\varphi \hat J_z}
     e^{-i\pi \hat J_x/2},
\end{equation}
which is equivalent to $U_{R_2}(\varphi)=\exp\{i\varphi \hat
J_y\}$. We obtain
%\begin{eqnarray}
%|\psi\rangle_{R_2}^{\rm even} &=\frac{1}{\sqrt{2}}
% &\left[e^{i\tfrac{\pi}{4}+i\tfrac{N\varphi}{2}}
% |\tfrac{\pi}{2},\tfrac{\pi}{2}\rangle\right.\nonumber\\
% &&\ \left.+(-1)^{\frac{N}{2}}e^{-i\frac{\pi}{4} -i\frac{N\varphi}{2}}
% |\tfrac{\pi}{2},\tfrac{3\pi}{2}\rangle\right]. \label{cat2ev}
%\end{eqnarray}
\begin{equation}
|\psi\rangle_{R_2}^{\rm even} =U_{R_2}(\varphi)
|\psi\rangle_{C_1}^{\rm even} = \frac{1}{\sqrt{2}}
 \left[e^{i\tfrac{\pi}{4}+i\tfrac{N\varphi}{2}}
 |\tfrac{\pi}{2},\tfrac{\pi}{2}\rangle
 +(-1)^{\frac{N}{2}}e^{-i\frac{\pi}{4} -i\frac{N\varphi}{2}}
 |\tfrac{\pi}{2},\tfrac{3\pi}{2}\rangle\right]. \label{cat2ev}
\end{equation}
Thus, the two state components have acquired opposite phase shifts
$\pm N\varphi/2$, proportional to $N$.

\begin{figure}[t]
\begin{center}
%\vspace{2cm}
\includegraphics[width=0.45\textwidth]{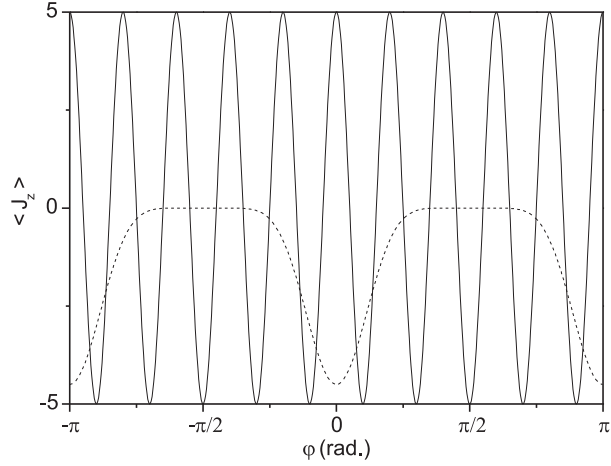}
%\framebox[0.8\columnwidth]{\huge FIGURE3}
\end{center}
 \caption{\label{fig:NevenOdd} Interferometric signal
$\medial {\hat J_z}$ after the entire Ramsey pulse sequence,
according to Eqs.~(\ref{evensig}) and (\ref{oddsig}). Shown are
the cases for even ($N=10$, solid line) and odd ($N=9$, dashed
line) atom number. The signals differ significantly in shape
depending on the parity of the atom number.}
\end{figure}

Then the atoms enter $C_2$, where they undergo the same
transformation as in $C_1$, $U_{C_2}=U_{C_1}$. It is easy to see
that the atomic state becomes:
\begin{equation}
|\psi\rangle_{C_2}^{\rm even}
=-\sin\bigl(\tfrac{N\varphi}{2}\bigr)|\tfrac{\pi}{2},\pi \rangle+
(-1)^{\tfrac{N}{2}}\cos\bigl(\tfrac{N\varphi}{2}\bigr)
|\tfrac{\pi}{2},0\rangle. \label{cat3ev}
\end{equation}
It is still an entangled atomic Schr\"odinger cat state, directed
along the $x$-axis, with $\varphi$-dependent amplitudes. Since we
always detect atomic energies, i.e., the number of atoms in
$|g\rangle$ or $|e\rangle$, we use a final $\pi/2$ pulse in zone
$R_3$ before the detection, $U_{R_3}=U_{R_1}$. The cat state of
Eq.~(\ref{cat3ev}) is rotated around the $y$-axis and put along
the $z$ direction, so that one obtains the final state
\begin{equation}
 |\psi\rangle_{\rm final}^{\rm even} =
 -\sin\bigl(\tfrac{N\varphi}{2}\bigr)|\theta=0
 \rangle
 + (-1)^{\tfrac{N}{2}}\cos\bigl(\tfrac{N\varphi}{2}\bigr)
|\theta=\pi\rangle. \label{cat4ev}
\end{equation}
%\begin{eqnarray}
% |\psi\rangle_{\rm final}^{\rm even}
% &=&-\sin\bigl(\tfrac{N\varphi}{2}\bigr)|\theta=0
% \rangle\nonumber\\
% &&+ (-1)^{\tfrac{N}{2}}\cos\bigl(\tfrac{N\varphi}{2}\bigr)
%|\theta=\pi\rangle. \label{cat4ev}
%\end{eqnarray}
A measurement of the expectation value of $\medial{\hat J_z}$ is
done by counting the number of atoms in $\ket{e}\,\, \hat=$
\mbox{$\ket{\theta = \pi}$} and $\ket{g}\hat = $
\mbox{$\ket{\theta = 0}$}. Using the corresponding operators $\hat
N_e = N/2+ \hat J_z$ and $\hat N_g = N/2 - \hat J_z$, we obtain
\begin{equation}\label{}
    \hat J_z = \frac{\hat N_e - \hat N_g}{2}.
\end{equation}
The expectation value of $\hat J_z$ in state $\ket{\psi}_{\rm
final}^{\rm even}$ is
\begin{equation}
\langle \hat J_z \rangle^{\rm even} = \frac{N}{2} \cos (N\varphi)
\label{evensig}.
\end{equation}
This signal displays Ramsey fringes with unit contrast oscillating
$N$ times faster than those corresponding to a single atom (see
Fig.~\ref{fig:NevenOdd}). To calculate the sensitivity of $\langle
\hat J_z \rangle$ to changes of $\varphi$, we evaluate
\cite{Leibfried2002}
\begin{equation}\label{eDeltaPhidef}
    \Delta \varphi = \frac{\Delta \hat J_z}{|\partial \medial{ \hat J_z}/ \partial
    \varphi|},
\end{equation}
where $\Delta \hat J_z =\sqrt{\medial{ \hat J_z^{\,2}}-\medial{
\hat J_z}^2}$ measures the fluctuations of the operator $\hat
J_z$. Eq.~(\ref{cat4ev}) yields $\medial{ \hat J_z^{\,2}}^{\rm
even}= N^2/4$ so that, using
Eqs.~(\ref{evensig})-(\ref{eDeltaPhidef}), one obtains $\Delta
\varphi= 1/N$, independent of $\varphi$. This means that, for an
even number of atoms, we obtain a $\sqrt{N}$ increase of
sensitivity with respect to a standard Ramsey experiment involving
$N$ uncorrelated atoms \cite{wine96} and thus achieve the
Heisenberg limit.

Grouping in a different way the unitary operations, we stress the
link between this scheme and a genuine Ramsey interferometry
experiment (see Fig.~\ref{fig:scheme}). We divide the sequence of
operations into three parts $U_{i}, (i=1,2,3)$. The first,
$U_{1}=\exp{\{-i \pi \hat J_x/2\}} U_{C_1} U_{R_1}$, consists of
the $\pi/2$ pulse in $R_1$, the dispersive interaction in $C_1$
and the first $\pi/2$ pulse in $R_2$, (see curly brackets in
Fig.~\ref{fig:scheme}). These three operations prepare the $N$
atoms, initially in the atomic coherent state $|\theta =0\rangle$,
in the maximally entangled superposition
$\left[e^{i\frac{\pi}{4}}|\theta=\pi\rangle+
(-1)^{\frac{N}{2}}e^{-i\frac{\pi}{4}}
|\theta=0\rangle\right]/\sqrt{2}$ (in the case of $N$ even). The
second part corresponds to the application of the tunable phase
$\varphi$ through the Stark shift pulse, $U_2=\exp\{i\varphi \hat
J_z\}$. Finally, the third part, $U_{3}= U_{R_3} U_{C_2} \exp{\{i
\pi \hat J_x/2 \}}$, consists of the second $\pi/2$ pulse in
$R_2$, the dispersive interaction in $C_2$ and the final $\pi/2$
pulse in $R_3$. The operations $U_1$ and $U_3$ are thus
``Super-Ramsey pulses'', corresponding to the $\pi/2$ pulses of a
classical Ramsey experiment.

In the preceding discussion we only considered the case of an even
atom number. For odd $N$ values, the state $|\psi\rangle_{R_2}$ at
the exit of $R_2$ becomes
%\begin{eqnarray}\label{psi_R2_odd}
%|\psi\rangle_{R_2}^{\rm odd}&
% =\dfrac{1}{\sqrt{2}}&\!\!\!\left[e^{i\tfrac{\pi}{4}}|\frac{\pi}{2}-\varphi,\pi\rangle\right.\nonumber\\
% &&\ \left.+ (-1)^{\frac{N-1}{2}}e^{-i\frac{\pi}{4}}
%|\tfrac{\pi}{2}+\varphi,0\rangle\right], \label{cat2od}
%\end{eqnarray}
\begin{equation}\label{psi_R2_odd}
|\psi\rangle_{R_2}^{\rm odd}
 =\dfrac{1}{\sqrt{2}}\left[e^{i\tfrac{\pi}{4}}|\frac{\pi}{2}-\varphi,\pi\rangle
 + (-1)^{\frac{N-1}{2}}e^{-i\tfrac{\pi}{4}}
|\tfrac{\pi}{2}+\varphi,0\rangle\right], \label{cat2od}
\end{equation}
for $|\varphi|\leq\pi/2$. For larger values of $|\varphi|$,
similar expressions can be derived. $|\psi\rangle_{R_2}^{\rm odd}$
is still a maximally entangled cat state, however oriented in the
$xz$ plane at an angle $\varphi$ with the $x$ axis. The Stark
shift pulse does not apply a relative phase on the state
components, but rather rotates them around the $z$ axis by an
angle $\varphi$. As a consequence, the final state of the $N$
atoms after $R_3$ is very different from that of
Eq.~(\ref{cat4ev}):
\begin{equation}
|\psi\rangle_{\rm final}^{\rm odd}
 =\dfrac{1}{2}\left[|\varphi,\pi\rangle +
 |\varphi,0\rangle
 -i(-1)^{\frac{N-1}{2}}\Bigl(
 |\pi-\varphi,\pi\rangle
   -|\pi-\varphi,0\rangle\Bigr)\right].
   \label{cat4od}
\end{equation}
%\begin{eqnarray}
% \!\!\!\!\!\! \!\!\!\!\!\!|\psi\rangle_{\rm final}^{\rm odd}
% &=&\dfrac{1}{2}\left[|\varphi,\pi\rangle +
% |\varphi,0\rangle\rule[0pt]{0pt}{12pt} \right.\nonumber\\
% &&\ \ \left.-i(-1)^{\frac{N-1}{2}}\Bigl(
% |\pi-\varphi,\pi\rangle
%   -|\pi-\varphi,0\rangle\Bigr)\rule[0pt]{0pt}{12pt}\right].
%   \label{cat4od}
%\end{eqnarray}
It is a superposition of \emph{four} coherent states, all in the
$xz$ plane (see Fig.~\ref{fig:blochSperes}). Using the properties
of atomic coherent states, it is possible to see that this state
yields an interferometric signal given by
\begin{equation}
\medial{\hat J_z}^{\rm odd} =-\frac{N}{2}\left (\cos
\varphi\right) ^{N-1}, \label{oddsig}
\end{equation}
very different from that for even $N$ (see
Fig.~\ref{fig:NevenOdd}).  It does not display any fast
oscillation, but only dips at $\varphi=0,\pm \pi$, having a width
inversely proportional to $\sqrt{N}$. Moreover Eq.~(\ref{cat4od})
yields $\medial{ \hat J_z^{\,2}}^{\rm odd}= \left(N^2\cos^2\varphi
+ N \sin^2\varphi\right)/4$ so that one obtains the following
phase sensitivity in the case of odd $N$
\begin{equation}
\label{oddsensi}
\Delta \varphi ^{\rm odd} = \frac{\sqrt{N^2[1-(\cos\varphi)^{2N-2}]-N(N-1)\sin^2\varphi}}{N
(N-1)|\sin\varphi||\cos\varphi|^{N-2}}.
\end{equation}
This expression is minimum at $\varphi=0,\pm \pi$, where $\Delta
\varphi ^{\rm odd}= 1/\sqrt{N}$, implying that, for an odd number
of atoms, the phase sensitivity is always worse, or at best equal,
to that of a standard Ramsey experiment involving $N$ uncorrelated
atoms \cite{wine96}.

\section{Interference signal averaged over the fluctuations of the atom number}

\label{sec:averaging}

\begin{figure}[tb]
%\includegraphics[width=0.40\textwidth]{nme10-d10om-v40-compno2.eps}
%\includegraphics[width=0.40\textwidth]{nme10-d2om-v180-oppno2.eps}
%\begin{center}
\centerline{ \includegraphics[width=0.9\textwidth]{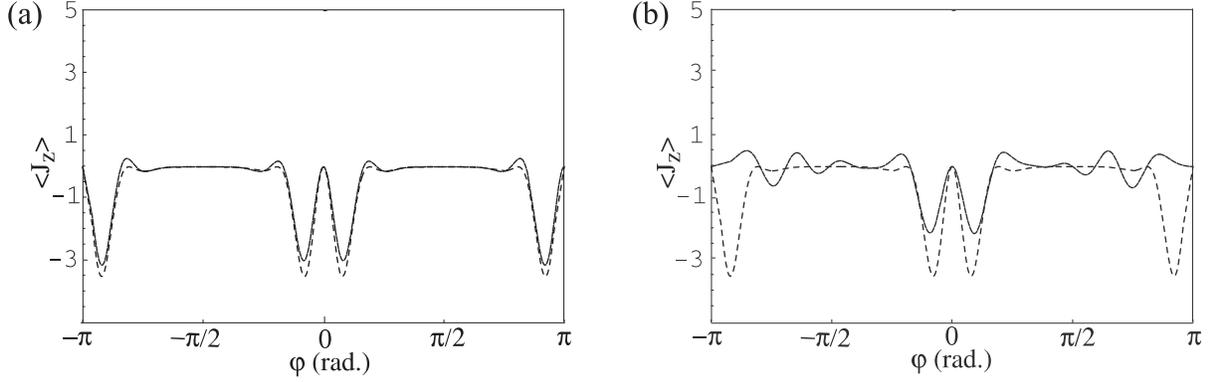}}
%\end{center}
\caption{\label{fig:disperno} Interferometric signal $\langle \hat
J_z \rangle$ for Poisson distributed number of atoms, with mean
number $\bar{N}=10$. The dashed lines show
Eq.~(\protect\ref{avesigno}), valid in the dispersive limit. The
full line is the numerical solution of the exact dynamics out of
the dispersive regime. (a) $\delta/\Omega_0 =9.27$, $v=40$~m/s.
(b) $\delta/\Omega_0 =2.06$, $v=180$~m/s (see text for the other
parameters).}
\end{figure}

An experimental implementation of the scheme described above will
be affected by the Poissonian fluctuations of the number of atoms
and a non-ideal detection efficiency of the detectors, which makes
it impossible to infer the parity of $N$. If we assume that we
prepare $N$ Rydberg atoms with a Poissonian probability
distribution with mean number $\bar{N}$ and average all detection
events, the resulting interferometric signal is given by the
average over the distribution of a signal given by
Eq.~(\ref{evensig}) for even $N$, and by Eq.~(\ref{oddsig}) for
odd $N$. The corresponding expression is evaluated analytically:
\begin{equation}
 \langle \hat J_z\rangle
=\frac{\bar{N}}{2}e^{-\bar{N}} \Bigl\{\sinh(u_\varphi)
\cos(v_\varphi)\cos\varphi -\cosh(u_\varphi)
 \left[1+\sin(v_\varphi)\sin\varphi\right]\Bigr\},
\label{avesigno}
\end{equation}
with $u_\varphi = \bar{N}\cos\varphi$, $v_\varphi =
\bar{N}\sin\varphi$.  The behavior of this signal for $\bar{N}=10$
is shown in Fig.~\ref{fig:disperno}(a) (dashed line), where it is
compared with the prediction of a numerical solution of the
dynamics driven by the exact Hamiltonian [Eq.~(\ref{tavcumm1})] in
the presence of cavity damping (full line). The fast oscillating
fringes of Eq.~(\ref{evensig}) except the central one at
$\varphi=0$ are degraded by the incoherent average over different
frequencies. In addition, the contribution of the experimental
runs with an odd atom number reduces the height of the central
fringe.

The agreement of Eq.~(\ref{avesigno}) with the numerical
simulation is good since we have chosen a parameter region
corresponding to the dispersive regime of large detunings. Using
the values of Ref.~\cite{rmp}, i.e., $\Omega_0=0.31 \times
10^6$~s$^{-1}$, $w=6$~mm, an atomic decay rate $\gamma_{\rm a} =
33.3$~s$^{-1}$ and a cavity decay rate $\gamma_{\rm c} =
10^{3}$~s$^{-1}$, Fig.~\ref{fig:disperno}(a) refers to
$\delta/\Omega_0=9.27$, i.e., $\delta =2.87 \times 10^6$~s$^{-1}$,
which, taking the condition of Eq.~(\ref{piphase}) into account,
implies an atomic velocity $v=40$~m/s. The resulting signal
$\langle \hat J_z\rangle$ is periodic with period  $\pi$. Its
contrast is smaller than 1/2 because of the signal suppression in
the wrong parity case. Despite that, this signal still scales with
the number of atoms as the Heisenberg limit, thanks to the peak at
$\varphi =0$, whose width is inversely proportional to $\bar{N}$,
as it can be seen from the approximate expression of
Eq.~(\ref{avesigno}) at small $\varphi$
\begin{equation}
\langle \hat J_z\rangle \simeq
-\frac{\bar{N}}{2}\exp\left(-\frac{\bar{N}\varphi ^2}{2}\right)
\sin^2\left[\frac{\varphi(\bar{N}+1)}{2}\right].
\label{avesignoapp}
\end{equation}
The exact phase sensitivity is provided by Eq.~(\ref{eDeltaPhidef}), whose explicit expression
in the case of Poisson-distributed number of atoms can be obtained from Eq.~(\ref{avesigno}) and the Poisson average
of the squared momentum
\begin{equation}
\langle \hat J_z^2\rangle =
\frac{\bar{N}}{4}
+ \frac{\bar{N}^2}{8}\left(1+\cos^2\varphi+e^{-2\bar{N}}
\sin^2\varphi\right).
\label{jzquapoi}
\end{equation}
The resulting expression is a cumbersome function of $\varphi$, achieving its minimum
at $|\varphi| \simeq \pi/(2\bar{N}+2)$, which also approximately corresponds to the points
where the absolute value of the slope of the
interferometric signal is maximum. At these phase shift values, $\Delta \varphi \simeq 2/\bar{N}$,
showing that at small values of $\varphi$ and in the dispersive regime,
our scheme reaches $1/2$ of the Heisenberg limit despite the presence
of a fluctuating number of atoms.

The low atomic velocity is hardly compatible with the state of the
art experiments \cite{rmp}. We thus consider the case of a more
realistic velocity $v=180$~m/s, for which we expect to have a
reasonable flux of atoms. In this case the $\pi/2$-pulse condition
of Eq.~(\ref{piphase}) imposes a smaller detuning $\delta/\Omega_0
=2.06$. The behavior of $\langle \hat J_z\rangle$ in this
non-dispersive regime is shown in Fig.~\ref{fig:disperno}(b) (full
line), where it is compared with the dispersive limit expression
of Eq.~(\ref{avesigno}) for $\bar{N}=10$ (dashed line). The
fringes at $\varphi= \pm \pi$ have been washed out, while the
central fringe at $\varphi=0$ is still visible, even though its
contrast is now significantly reduced. The corresponding
resolution in the estimation of $\varphi$ is now decreased and one has
$\Delta \varphi \simeq 3.3/\bar{N}$.

\section{Modified scheme with inversion of time evolution}
\label{sec:inversion}
\begin{figure}[b]
\begin{center}
\includegraphics[width=0.45\textwidth]{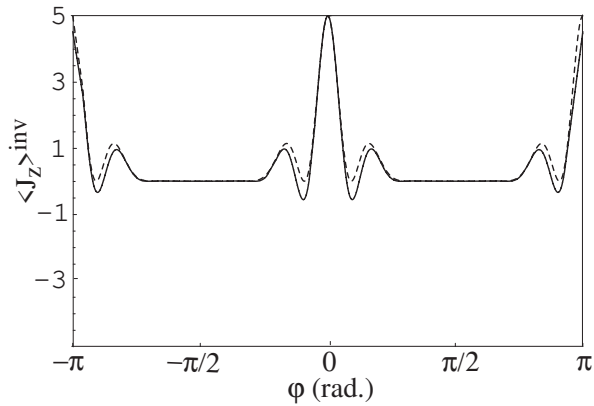}
\end{center}
\caption{\label{dispersi40} Interferometric signal $\langle \hat
J_z\rangle^{\rm inv}$ in the presence of the ``inversion'' of time
evolution in $C_2$, for Poisson-distributed number of atoms with
mean number $\bar{N}=10$. The dashed line refers to
Eq.~(\protect\ref{avesig}), which has been derived assuming the
dispersive Hamiltonian of Eq.~(\ref{disp2}), while the full line
refers to the numerical solution of the exact dynamics in the case
of large detuning, $\delta/\Omega_0 =9.27$ and $v=40$~m/s (see
text for the other parameters).}
\end{figure}

\begin{figure*}[ptbh]
\begin{center}
\includegraphics[width=0.8\textwidth]{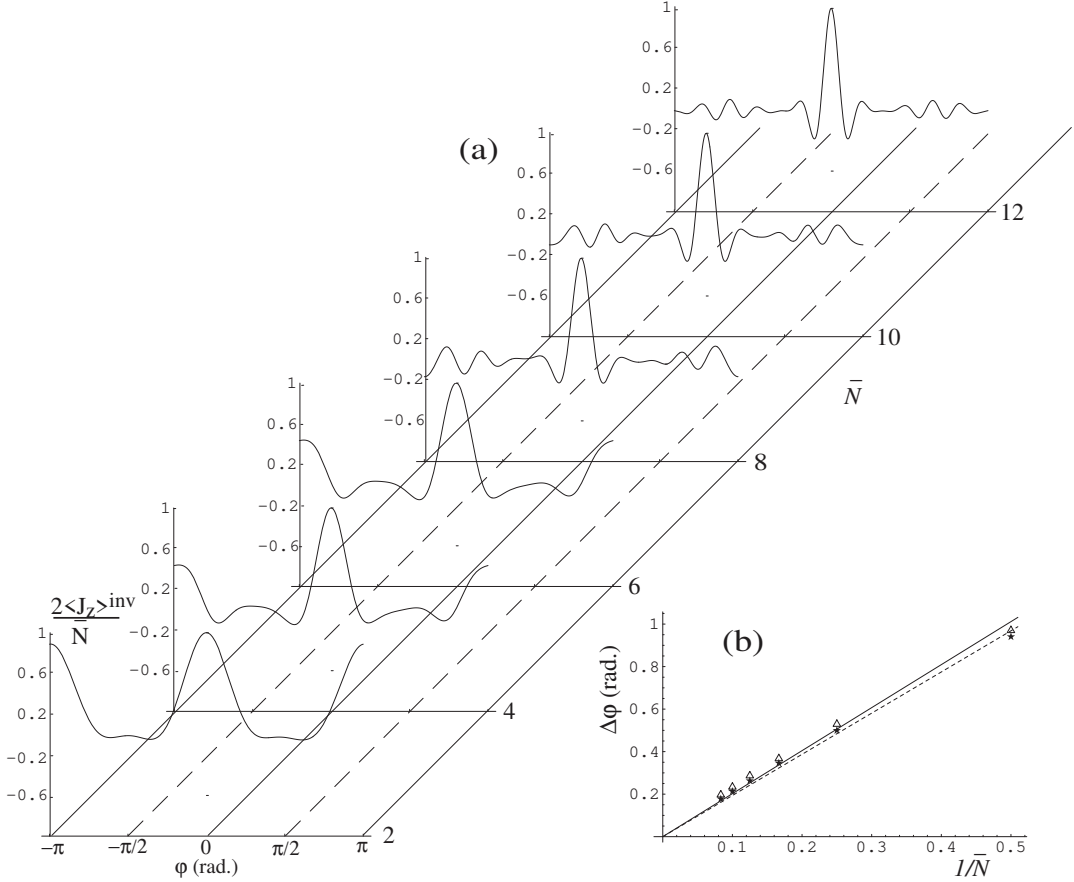}
\end{center}
\caption{\label{tutti180} (a) Normalized interferometric signal
$\langle 2\hat J_z\rangle^{\rm inv}/\bar{N}$ in the presence of
the inversion in $C_2$, for Poisson-distributed number of atoms,
with mean number $\bar{N} = 2,4,6,8,10,12$. The curves show the
numerical solution of the exact dynamics in the non-dispersive
regime, with $\delta/\Omega_0 =2.06$ and $v=180$~m/s. (b) Phase
uncertainty $\Delta \varphi$ as defined in
Eq.~(\protect\ref{eDeltaPhidef}) for the same parameter values,
versus $1/\bar{N}$. Stars refer to the minimum phase uncertainty
and the dotted line is the least-squares linear fit of the data,
while triangles refer to the phase uncertainty at the point of
maximum absolute slope of the signal and the full line is the
least-squares linear fit of the corresponding data.}
\end{figure*}

As shown in Fig.~\ref{fig:disperno}(b), the contrast of the
interference fringes degrades as soon as we leave the dispersive
limit of slow velocities and we consider the fluctuations of the
number of atoms. However, it is possible to significantly improve
the scheme if we avoid the suppression of the signal at $\varphi
=0$ in the odd $N$ case (see Eq.~(\ref{oddsig})) which is
responsible for the fact that the signal is practically never
larger than zero. This can be achieved if the time
evolution in $C_2$ is ``inverted'', i.e., $\hat H_{I,2}(t) \to
-\hat H_{I,2}(t)$. In the dispersive limit, this inversion can be
achieved by changing only the sign of the detuning $\delta_2$
through an appropriate Stark shift of the atomic levels. In such a
case at $\varphi=0$ nothing happens in $R_2$, and the two
interactions in $C_1$ and $C_2$ cancel each other. The atoms only
undergo the two $\pi/2$ pulses in $R_1$ and $R_3$ which rotate the
state from $|\theta=0\rangle $ to $|\theta=\pi\rangle$. The final
result is $\langle \hat J_z \rangle = N/2$ ($N_e=N$, $N_g=0$) at
$\varphi =0$ either in the even and in the odd case. Within the
dispersive limit, we derive the state of the $N$ atoms in the
presence of this time evolution inversion. Nothing changes for
even $N$, i.e., Eqs.~(\ref{cat3ev})-(\ref{evensig}) remain valid,
while in the odd $N$ case the final state is now given by
\begin{equation}
|\psi\rangle_{\rm final}^{\rm odd, inv}
 = \frac{1}{2}\biggl[|\varphi,0\rangle-
 |\varphi,\pi\rangle
 -i(-1)^{\frac{N-1}{2}}\Bigl( |\pi-\varphi,\pi\rangle
+|\pi-\varphi,0\rangle\Bigr)\biggr],
 \label{cat4odinv}
\end{equation}
yielding an interferometric signal
\begin{equation}
\langle \hat J_z \rangle^{\rm odd,inv} =\frac{N}{2}\left(\cos
\varphi\right)^{N-1} \label{oddsig2}.
\end{equation}
Using Eq.~(\ref{oddsig2}), the resulting Poisson-averaged
interferometric signal can be obtained by changing only the sign
of the term $\cosh\left(u_\varphi\right)$ in the
expression of Eq.~(\ref{avesigno}), so that
\begin{equation}
\langle \hat J_z \rangle ^{\rm inv}
 = \frac{\bar{N}}{2} e^{-\bar{N}}
  \Big[\sinh(u_\varphi)\cos(v_\varphi)\cos\varphi
% \nonumber \\
 +\cosh(u_\varphi)
 \left[1-\sin(v_\varphi)\sin\varphi\right]\Big],
\label{avesig}
\end{equation}
where again $u_\varphi = \bar{N}\cos\varphi$, $v_\varphi =
\bar{N}\sin\varphi$. The behavior of this expression for
$\bar{N}=10$ is shown in Fig.~\ref{dispersi40} (dashed line),
where it is compared with the corresponding value of $\langle \hat
J_z \rangle ^{\rm inv} $ obtained from the numerical solution of
the dynamics driven by the exact Hamiltonian of
Eq.~(\ref{tavcumm1}) and in the presence of cavity damping (full
line) for the same parameters chosen in
Fig.~\ref{fig:disperno}(a).

The signal is again periodic with period equal to $\pi$, but it
now shows clear peaks at $\varphi=0,\pm \pi$, where it reaches the
maximum value $\bar{N}/2$. The width of the peaks is again
inversely proportional to $\bar{N}$, because Eq.~(\ref{avesig})
can be approximated at small $\varphi$ as
\begin{equation}
\langle \hat J_z\rangle ^{\rm inv} \simeq \frac{\bar{N}}{2}
\exp\left(-\frac{\bar{N}\varphi ^2}{2}\right)
\cos^2\left[\frac{\varphi(\bar{N}+1)}{2}\right]. \label{avesigapp}
\end{equation}
The explicit expression of the phase sensitivity of Eq.~(\ref{eDeltaPhidef})
can be obtained from Eq.~(\ref{avesig}) and from Eq.~(\ref{jzquapoi}) for $\langle \hat J_z^2\rangle$,
which is valid also in the presence of the inversion.
The resulting expression is again a cumbersome function of $\varphi$, now achieving its minimum
at $|\varphi| \simeq \pi/(4\bar{N}+4)$, where $\Delta \varphi \simeq 1.45/\bar{N}$.
Differently from the preceding scheme without inversion, the point of minimum phase uncertainty does not coincide with
the point of maximum slope of the signal $\langle \hat J_z\rangle ^{\rm inv}$, which is again approximately equal to
$|\varphi| \simeq \pi/(2\bar{N}+2)$, as in the scheme with no inversion of time evolution of the preceding Section.
At these values of $\varphi$, the phase uncertainty is larger, but still scales as the Heisenberg limit
with the number of atoms, because it is $\Delta \varphi \simeq 1.65/\bar{N}$.

The inversion of the time evolution in $C_2$ becomes of crucial
importance in the non-dispersive regime. In this case, one also
has to change the sign of $\hat J_{\pm}$ in addition to reversing
the sign of $\delta_2$ in order to get $\hat H_{I,2}(t) \to -\hat
H_{I,2}(t)$. This second inversion can be realized by applying two
opposite $\pi$ Stark shifts. A first one, described by the
operator $\exp\{i\pi \hat J_z\}$, must be applied in $R_2$ just
before the entrance in $C_2$, while the second, described by the
operator $\exp\{-i\pi \hat J_z\}$, must be applied in $R_3$ soon
after the exit from $C_2$. The new unitary operators for the
Ramsey zones $R_2$ and $R_3$ then read:
\begin{eqnarray}\label{U2bis}
    U'_{R_2}(\varphi) &=& e^{i \pi \hat J_z} U_{R_2}(\varphi) ,\\
    U'_{R_3} &=& U_{R_3}e^{-i \pi \hat J_z} .
\end{eqnarray}
As a result of the inversion, the narrow central interference peak
at $\varphi=0$ becomes much more robust and it survives even
outside the dispersive regime of slow velocities. This behavior of
$\langle \hat J_z \rangle^{\rm inv}$ is shown in
Fig.~\ref{tutti180}, for different values of $\bar{N}$, while in
Fig.~\ref{tutti180}(b) the dependency $\Delta\varphi \simeq 2.0
/\bar{N}$ is clearly visible. This fact proves that
thanks to inversion in $C_2$, we are able to reach about one half of the
ultimate, Heisenberg-limited, resolution for the estimate of
$\varphi$, even in the non-dispersive regime of fast atoms.

Notice that in the non-dispersive regime the dynamics is no more
described by the Hamiltonian of Eq.~(\ref{disp2}) and therefore
the atomic phase shift is no more given by Eq.~(\ref{piphase}).
Nonetheless, we have still used Eq.~(\ref{piphase}) for the choice
of the value of $\delta$ at a given velocity $v$, because it
turned out in the numerical simulation to give the best results.
When fast atoms are used ($v=180$~m/s), the atomic state after the
interactions in $C_1$ and $C_2$ is rather different from that of
the dispersive limit [Eqs.~(\ref{cat1ev}) and (\ref{cat3ev})].
However, the interference experiment in the presence of the
inversion in $C_2$ still approaches the Heisenberg limit.

%{\bf Experimental limitations}

 The main experimental limitations affecting this scheme are
cavity damping, the spontaneous emission of the atoms and the
non-unit detection efficiency. Cavity damping is not a serious
limitation because the cavities play a passive role in the
experiment, only allowing the virtual exchange of excitations
between the atoms. The cavities are initially in the vacuum state
and can get photons from the excited Rydberg atoms only close to
resonance. Cavity decay starts to significantly affect the
generated maximally entangled GHZ atomic state only when
$\gamma_{\rm c} \simeq \delta \simeq \Omega_0\sqrt{N}$. This is
not a strong condition on $\gamma_{\rm c}$ because typically
$\delta \simeq 10^6$ s$^{-1}$. Thus, a moderate quality factor of
$Q = \omega_{eg}/\gamma_{\rm c}> 10^{6}$ would be sufficient,
compared to experimentally attainable values of $Q > 10^8$.

Also atomic spontaneous emission does not represent a serious
problem, thanks to the use of circular Rydberg levels with
lifetimes of the order of $30$ ms, even though it puts a practical
upper bound on the maximum number of atoms $N$. Spontaneous
emission implies a decoherence timescale of the maximally
entangled atomic GHZ states of the order of $30/N$ ms. This
decoherence time has to be longer than the time of flight through
the apparatus which is of the order of $1$ ms. Therefore it is
reasonable to perform the experiment with up to 20 atoms.

To consider a non-ideal detection efficiency, we assume that the
detector is characterized by the same quantum efficiency $\eta$
for detecting atoms in $e$ or $g$, that there are no dark counts,
and we neglect the possibility of a wrong state detection ($e
\leftrightarrow g$), the phase uncertainty simply reads $\Delta
\varphi' =\eta^{-1} \Delta \varphi$. Typical detection
efficiencies attainable in a cavity QED experiment with Rydberg
atoms are between 80\% and 100\% \cite{Maioli}.

\section{Conditioning to the number of detected atoms}

\label{sec:conditioning}

In the preceding schemes, we did not use any information on the
number of detected atoms. We now show that we can improve the
phase sensitivity by means of an appropriate data processing of
the signals conditioned to the number of detected atoms. If one
post-selects only the data corresponding to a given number of
detected atoms $N_d = N_e^{\rm det} + N_g^{\rm det}$, in the ideal
case of a detector with quantum efficiency $\eta=1$, one would get
an interference signal coinciding with the ideal one
[Eq.~(\ref{evensig})] with $N=N_d$ (even), thanks to the
conditional generation of a maximally entangled GHZ state of $N_d$
atoms. If instead the detector is not perfect, the interference
signal is the result of an average over many experimental runs in
which the actual number of atoms is equal or larger than $N_d$. As
a consequence, the contrast of the interference fringes rapidly
worsens for a decreasing detection efficiency $\eta$. This
post-selection strategy is in a certain sense opposite to the one
considered in the preceding sections, where we have averaged the
interferometric signal over all the runs, regardless the value of
the number of detected atoms. In the post-selection case, one
gives zero weight to the signal conditioned to a number of
detected atoms $N_d$ different from the selected one. The
Poissonian-averaged signal of the preceding section instead
corresponds to keep all the data and to give the same weight to
the conditioned signals.

It is evident that the post-selection strategy is not optimal for
achieving the maximum phase sensitivity, because it implies
wasting most of the resources, i.e., all the atoms of the runs
with a number of detected atoms different from the selected one
$N_d$. %In principle one could achieve a lower phase uncertainty
%$\Delta \varphi$ by using the total number of employed atoms
%$N_{\mathrm{tot}}$ in a standard Ramsey interferometry experiment
%with uncorrelated atoms, where $\Delta \varphi =
%N_{\mathrm{tot}}^{-1/2}$.
In this experiment, the best strategy for achieving the maximum
phase sensitivity is to keep all the data, however grouping them
into different sets according to the corresponding number of
detected atoms $N_d$, and then give appropriate weights $w(N_d)$
to the signal conditioned to the detection of $N_d$ atoms,
$\langle \hat{J}_z(N_d)\rangle^{cond}$. This corresponds to
consider the following interferometric signal
\begin{equation}
\langle \hat{J}_z \rangle^{w}= \sum_{N_d=0}^{\infty}P(N_d) w(N_d)\langle \hat{J}_z(N_d)\rangle^{cond}, \label{weisig}
\end{equation}
where $P(N_d)$ is the probability of detecting $N_d$ atoms which, due to the assumptions
made above, is given by
\begin{equation}
P(N_d)=e^{-\eta \bar{N}}\frac{\left(\eta \bar{N}\right)^{N_d}}{N_d !}.
\label{probcond}
\end{equation}
The optimized signal corresponds to take the weights $w^{\rm
opt}(N_d)$ which, for each $\varphi$, minimize the phase
uncertainty $\Delta \varphi$ of Eq.~(\ref{eDeltaPhidef}), which
for the signal of Eq.~(\ref{weisig}) has the explicit form
%\begin{widetext}
\begin{equation}
\Delta \varphi = \frac{\left[\sum_{N_d=0}^{\infty}P(N_d) w(N_d)^2\langle \hat{J}_z^2(N_d)\rangle^{cond}-
\left(\sum_{N_d=0}^{\infty}P(N_d) w(N_d)\langle \hat{J}_z(N_d)\rangle^{cond}\right)^2\right]^{1/2}}
{\left|\sum_{N_d=0}^{\infty}P(N_d) w(N_d)\partial \langle \hat{J}_z(N_d)\rangle^{cond}/\partial \varphi \right|}. \label{explfase}
\end{equation}
%\end{widetext}
The Poisson-averaged signal of the preceding sections corresponds to the particular case $w(N_d)=1$ $\forall N_d$ in Eq.~(\ref{weisig}).

\begin{figure}[t]
\begin{center}
\includegraphics[width=0.95\textwidth]{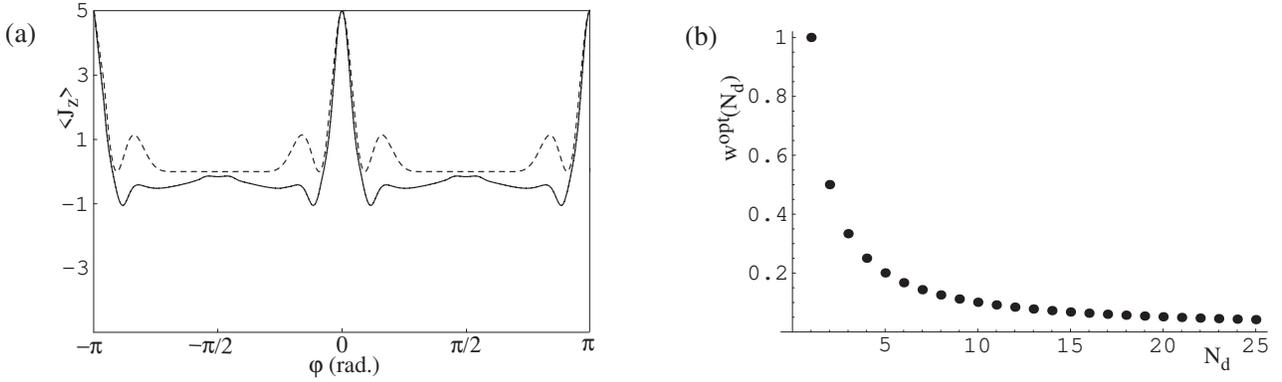}
\end{center}
\caption{\label{fig:conditioning} (a) Optimal interferometric
signal $\langle \hat J_z \rangle^{\rm opt}$ of
Eq.~(\ref{weisigopt}) (full line) versus the phase shift applied in $R_2$,
$\varphi$. The optimal weights are determined by the condition of
minimum phase uncertainty at each $\varphi$. The dashed line refers to the
Poisson-averaged signal of
Eq.~(\protect\ref{avesig}), corresponding to take equal weights. (b) Weights  $w^{\rm
opt}(N_d)$ of the optimal signal versus $N_d$ at the point of the maximum
phase sensitivity. The plots refer to $\eta=0.8$ and
$\bar{N}=12.5$ and to the dispersive limit of large detuning,
where Eqs.~(\protect\ref{avesigcond}) and
(\protect\ref{avesigcond2}) apply.}
\end{figure}

We illustrate this optimization strategy by applying it to the experiment with the inversion of time evolution in $C_2$,
in the dispersive limit of large detuning.
In this limit, using Eqs.~(\ref{evensig}) and (\ref{oddsig2}), it
is possible to find, after long but straightforward calculations, that the conditional signal is given by
\begin{eqnarray}\label{avesigcond}
&&\langle \hat J_z(N_d)\rangle^{\rm cond} = \frac{N_d}{4}
e^{-\bar{N}_l}\left\{e^{\bar{N}_l\cos\varphi} \left[\cos
(N_d\varphi+\bar{N}_l\sin\varphi) + (\cos\varphi)^{N_d-1}\right]  \right.\nonumber \\
&& \hspace{4.2cm}\left. +(-1)^{N_d} e^{-\bar{N}_l\cos\varphi}
\left[\cos (N_d\varphi-\bar{N}_l\sin\varphi) -
(\cos\varphi)^{N_d-1}\right]\right\},
\end{eqnarray}
where $\bar{N}_l = \bar{N}(1-\eta)$, is the mean number of atoms
lost by the detector.
Moreover in the dispersive limit one also finds
\begin{equation}\label{avesigcond2}
\langle \hat{J}_z^2(N_d)\rangle^{\rm cond}
= \frac{N_d^2}{4}+\frac{N_d-N_d^2}{8}\sin ^2 \varphi \left[1-(-1)^{N_d}e^{-\bar{N}_l}\right].
\end{equation}
These two latter expressions are then inserted into the expression
for the phase uncertainty $\Delta \varphi$ of
Eq.~(\ref{explfase}), which is then minimized with respect to the
weights $w(N_d)$. The corresponding optimal weights $w^{\rm
opt}(N_d)$ depend upon the phase shift applied in $R_2$,
$\varphi$, due to the dependence of $\Delta \varphi$ upon
$\varphi$, and when inserted into Eq.~(\ref{weisig}), one gets the
optimal interference signal
\begin{equation}
\langle \hat{J}_z \rangle^{\rm opt}= \sum_{N_d=0}^{\infty}P(N_d)
w^{\rm opt}(N_d)\langle \hat{J}_z(N_d)\rangle^{cond}.
\label{weisigopt}
\end{equation}
This optimized signal is shown in Fig.~\ref{fig:conditioning}(a) (full line),
which refers to the case $\eta =0.8$ and $\bar{N}=12.5$. The
corresponding minimum phase uncertainty is a function of
$\varphi$, and its minimum value $\Delta \varphi_{\rm min}$ is
closer to the ideal Heisenberg limit: $\Delta \varphi_{\rm min}
\simeq 1.3/\eta\bar{N}$. Comparing with the result of the
preceding section, we see that optimizing the information provided
by conditioning on the number of detected atoms yields an improved
phase sensitivity, since the corresponding minimum phase
uncertainty of the Poisson-averaged case scales as $\Delta
\varphi_{\rm min} \simeq 1.45/\eta\bar{N}$. As it happens in the
Poisson-averaged signal, the best phase sensitivity is achieved at
a phase shift $\varphi$ \emph{smaller} than that corresponding to
the maximum slope of the optimal signal $\langle \hat{J}_z
\rangle^{\rm opt}$, which is approximately equal to $|\varphi|
\simeq \pi/(3\eta \bar{N})$. At this value of the phase shift, the
phase uncertainty still scales as the Heisenberg limit, but is
larger and it is roughly given by $\Delta \varphi \simeq
1.5/\eta\bar{N}$.

It is also interesting to see the behavior of the optimal weights
$w^{\rm opt}(N_d)$ at the phase shift $\varphi$ corresponding to
the maximum phase sensitivity, which are shown in
Fig.~\ref{fig:conditioning}(b), again for the case $\eta =0.8$ and
$\bar{N}=12.5$. One can notice that they monotonically increase
for decreasing $N_d$ and this is due to the fact that the optimal
weights tends to be inversely proportional to $\langle
\hat{J}_z^2(N_d)\rangle^{\rm cond}$. From the analysis of Section
II, one could have expected an oscillating behavior of $w^{\rm
opt}(N_d)$, with maxima corresponding to $N_d$ even and minima
corresponding to $N_d$ odd. However it can be seen that this
happens only in the limit $\eta \to 1$, when the probability to
detect a wrong parity of the number of atoms becomes negligible.
Only in this limit, one can safely discriminate between an even
and an odd number of atoms and suppress the contribution of the
runs with an odd number of atoms.

%\begin{figure}[t]
%\begin{center}
%\includegraphics[width=0.45\textwidth]{figure8.eps}
%\end{center}
%\caption{\label{fig:weights} }
%\end{figure}

\section{Conclusions}\label{sec:conclusions}

We have presented a Ramsey-like interference experiment for a
cavity-QED system, able to reach the ultimate Heisenberg limit for
the estimation of an atomic phase-shift. We considered a system of
$N$ two-level Rydberg atoms successively crossing two microwave
cavities. A dispersive atom cavity interaction is able to generate
an atomic Schr\"odinger cat, i.e., a superposition of two atomic
coherent states, which represents a maximally entangled state of
the $N$ atoms. Using this state, we designed an interference
experiment yielding fringes $N$ times narrower than those one
would have obtained if $N$ disentangled atoms were used.

As discussed in Sec.~\ref{sec:averaging}, the most important
limitations affecting the the experiment are the fluctuations of
the number of Rydberg atoms, and the non-unit detection
efficiency. Despite these limitations, one can approach
Heisenberg-limited sensitivity, because a
narrow central fringe, with a width inversely proportional to
$\bar{N}$, survives even when averaged over the fluctuations of
the number of detected atoms.

Finally we have considered a conditional scheme in which we
post-select only the events with a fixed number of detected atoms.
In such a case, the best strategy for achieving the maximum phase
sensitivity is to keep all the data, however grouping them into
different sets according to the corresponding number of detected
atoms $N_d$, and then give appropriate weights $w(N_d)$ to the
signal conditioned to the detection of $N_d$ atoms, $\langle
\hat{J}_z(N_d)\rangle^{cond}$. One can then determine the optimal
weights maximizing the phase sensitivity and the corresponding
optimized signal closely approaches the ultimate Heisenberg limit,
because the minimum phase uncertainty is $\Delta \varphi_{\rm min}
\simeq 1.3/\eta\bar{N}$.

The proposed experiment should allow for the first time to achieve
the Heisenberg limit for spectroscopy with a larger number of
atoms compared to the experiments realized so far with entangled
photons or ions. Despite this application in spectroscopy,
characterizing the maximally entangled Schr\"odinger cat state by
e.g. state tomography and monitoring its decoherence would be a
very interesting perspective.

\section{Acknowledgements}
We acknowledge the funding by the city of Paris (D.V.), by a
Marie-Curie fellowship of the European Community (S.K.) and by the
EU under the IP projects ``QGATES'' and ``SCALA''.

\end{document}